\documentclass[%
 reprint,
superscriptaddress,
 amsmath,amssymb,
 aps,
floatfix,
]{revtex4-1}

\usepackage{graphicx}
\usepackage{dcolumn}
\usepackage{bm}
\usepackage{hyperref}

\begin{document}

\preprint{APS/123-QED}

\title{Evidence against the Efimov effect in $^{12}\mathrm{C}$ from spectroscopy and astrophysics}
		\author{J.~Bishop}
		\email{jackbishop@tamu.edu}

		\affiliation{Cyclotron Institute, Texas A\&M University, College Station, TX 77843, USA}
		\affiliation{Department of Physics \& Astronomy, Texas A\&M University, College Station, TX 77843, USA}
		\author{G.V.~Rogachev}
		\affiliation{Cyclotron Institute, Texas A\&M University, College Station, TX 77843, USA}
		\affiliation{Department of Physics \& Astronomy, Texas A\&M University, College Station, TX 77843, USA}
		\affiliation{Nuclear Solutions Institute, Texas A\&M University, College Station, TX 77843, USA}
		\author{S.~Ahn}
				\affiliation{Cyclotron Institute, Texas A\&M University, College Station, TX 77843, USA}
		\affiliation{Center for Exotic Nuclear Studies, Institute for Basic Science, Daejeon, 34126, Korea}
		\author{E.~Aboud}
				\affiliation{Cyclotron Institute, Texas A\&M University, College Station, TX 77843, USA}
		\affiliation{Department of Physics \& Astronomy, Texas A\&M University, College Station, TX 77843, USA}
		\author{M.~Barbui}
				\affiliation{Cyclotron Institute, Texas A\&M University, College Station, TX 77843, USA}
		\author{A.~Bosh}
				\affiliation{Cyclotron Institute, Texas A\&M University, College Station, TX 77843, USA}
		\affiliation{Department of Physics \& Astronomy, Texas A\&M University, College Station, TX 77843, USA}
		\author{ J.~Hooker} 
				\affiliation{Department of Physics \& Astronomy, University of Tennessee, Knoxville, TN 37996, USA}
		\author{C.~Hunt}
				\affiliation{Cyclotron Institute, Texas A\&M University, College Station, TX 77843, USA}
		\affiliation{Department of Physics \& Astronomy, Texas A\&M University, College Station, TX 77843, USA}
		\author{H.~Jayatissa}
				\affiliation{Physics Division, Argonne National Laboratory, Argonne, IL 60439, USA}
		\author{E.~Koshchiy}
				\affiliation{Cyclotron Institute, Texas A\&M University, College Station, TX 77843, USA}
		\author{R.~Malecek}
		\affiliation{{Department of Physics and Astronomy, Louisiana State University, Baton Rouge, LA 70803, USA} }
		\author{S.T.~Marley}
				\affiliation{{Department of Physics and Astronomy, Louisiana State University, Baton Rouge, LA 70803, USA} }
		\author{M.~Munch}
				\affiliation{Department of Physics and Astronomy, Aarhus University, DK-8000 Aarhus C, Denmark}
		\author{E.C.~Pollacco}
		\affiliation{{IRFU, CEA, Saclay, Gif-Sur-Ivette, France}}
		\author{C.D.~Pruitt}
				\affiliation{Department of Chemistry, Washington University, St. Louis, MO 63130, USA}
				\affiliation{Lawrence Livermore National Laboratory, Livermore, CA 94550, USA}
		\author{B.T.~Roeder}
						\affiliation{Cyclotron Institute, Texas A\&M University, College Station, TX 77843, USA}

		\author{A. Saastamoinen}
						\affiliation{Cyclotron Institute, Texas A\&M University, College Station, TX 77843, USA}

		\author{L.G.~Sobotka}
		\affiliation{Department of Chemistry, Washington University, St. Louis, MO 63130, USA}
		\author{S.~Upadhyayula}
				\affiliation{TRIUMF, 4004 Westbrook Mall, Vancouver, British Columbia V6T 2A3, Canada}

\date{\today}

\begin{abstract}
\begin{description}
\item[Background]
The Efimov effect is a universal phenomenon in physics whereby three-body systems are stabilized via the interaction of an unbound two-body sub-systems. A hypothetical state in $^{12}\mathrm{C}$ at 7.458 MeV excitation energy, comprising of a loose structure of three $\alpha$-particles in mutual two-body resonance, has been suggested in the literature to correspond to an Efimov state in nuclear physics. The existence of such a state has not been demonstrated experimentally.
\item[Purpose]
Using a combination of $\gamma$-spectroscopy, charged-particle spectroscopy and astrophysical rate calculations, strict limits on the existence of such a state have been established here.
\item[Method]
Using the combined data sets from two recent experiments, one with the TexAT TPC to measure $\alpha$-decay and the other with Gammasphere to measure $\gamma$-decay of states in $^{12}\mathrm{C}$ populated by $^{12}\mathrm{N}$ and $^{12}\mathrm{B}$ $\beta$-decay respectively, we achieve high sensitivity to states in close-proximity to the $\alpha$-threshold in $^{12}\mathrm{C}$.
\item[Results]
No evidence of a state at 7.458 MeV is seen in either data set. Using a likelihood method, the 95\% C.L. $\gamma$-decay branching ratio is determined as a function of the $\beta$-decay feeding strength relative to the Hoyle state. In parallel, calculations of the triple-alpha reaction rate show the inclusion of the Efimov corresponds to a large increase in the reaction rate around $5 \times 10^{7}$ K.
\item[Conclusion]
From decay spectroscopy - at the 95\% C.L., the Efimov state cannot exist at 7.458 MeV with any $\gamma$-decay branching ratio unless the $\beta$-strength is less than 0.7\% of the Hoyle state. This limit is evaluated for a range of different excitation energies and the results are not favorable for existence of the hypothetical Efimov state in $^{12}\mathrm{C}$. Furthermore, the triple-alpha reaction rate with the inclusion of a state between 7.43 and 7.53 MeV exceeds the rate required for stars to undergo the red giant phase.
\end{description}
\end{abstract}

\pacs{Valid PACS appear here}
\maketitle

\section{\label{sec:Introduction}Introduction}
The Efimov effect is a universal quantum phenomena present in several areas of Physics. The effect is observed for three-body systems that are comprised of subsystems where the sub-unit two-body systems are unbound but have a large s-wave scattering length. Vitaly Efimov found that under these conditions the long-range three-body attraction arises and that this attraction can support a family of three-body states. A detailed review of Efimov physics and experimental evidence for it can be found in Ref. \cite{EfimovCoulomb}. The classical interpretation is that the three-body force binds the system via the `shuttling' back and forth of one of the particles. This then creates an infinite series of states given by a universal scaling law. While predicted 50 years ago, this phenomenon took 35 years to observe experimentally. The first clear evidence for Efimov effect was reported for the system of ultracold gas of caesium atoms in an external magnetic field \cite{Kra2006}. In nuclear systems, which were the original focus of Efimov's investigation, the situation is more complicated. In principle, a $J^{\pi}=0^{+}$ (corresponding to L=0) 3-$\alpha$ state in $^{12}\mathrm{C}$, where the 2-$\alpha$ sub-systems are unbound but form a long-lived resonant state, can be seen as Efimov trimer. In his original paper Efimov argued that the 0$^+$ excited state at 7.65 MeV in $^{12}\mathrm{C}$ (Hoyle state) possibly originates due to this interesting three-body quantum phenomena which we now call the Efimov effect \cite{Efimov1970}. Microscopic three-body continuum calculations that utilize phenomenological $\alpha$-$\alpha$ potential \cite{Sun2015} indicate that the centrifugal (three-body) potential is small compared to the nuclear and Coulomb potentials and as a result the Hoyle state is probably not related to the Efimov effect. Faddeev calculations for the 3$\alpha$ system have previously been performed and do not predict an additional near-threshold $0^{+}$ that would correspond to an Efimov state \cite{Faddeev}. More recently, existence of an Efimov state in $^{12}\mathrm{C}$ at an excitation energy that corresponds to a mutual $^{8}\mathrm{Be}$(g.s.) resonance for all three $\alpha$-particles was suggested in \cite{Zhe2018,Zheng_2020}. This excitation energy is given, in units of MeV, by the following simple relation which takes into account that the narrow resonance of $^{8}\mathrm{Be}$(g.s.) is unbound by 91.84 keV with respect to decay to two $\alpha$-particles:

\begin{align}
E = \frac{2}{3} \sum^{3}_{i}E_{rel_{i}} - Q = 2\times0.09184 + 7.2747 = 7.458.
\end{align}
\par

The main goal of this paper is to search for any evidence of this hypothetical state or, if not, place experimental limits on its existence. We utilize the results of two recent experiments. The first one is the study that is sensitive to three $\alpha$-particles decay channel of the near $\alpha$-threshold excited states in $^{12}\mathrm{C}$ populated in the $\beta^{+}$ decay of $^{12}\mathrm{N}$ \cite{NIMJEB,Bis2020}. The second is the $\gamma$-spectroscopy study of states in $^{12}\mathrm{C}$ populated in $\beta^{-}$ decay of $^{12}\mathrm{B}$ \cite{Munch}. Combining these two data sets we demonstrate that there is no evidence for a resonance at or near 7.458 MeV. Stringent experimental limits on its existence have been established. 
We also examine the astrophysical implications of such a state existing below the Hoyle state using a simplistic model to calculate the triple-alpha reaction rate and demonstrate the incompatibility of this result with current astrophysical observations.

\section{Experimental limits of an Efimov state}

There has not been a large amount of experimental activity to investigate if there is a resonance in $^{12}\mathrm{C}$  below the Hoyle state. One investigation claiming to observe an Efimov state in a heavy-ion Zn+Zn/Ni+Ni collisions at 35 MeV/u was recently published \cite{Aldo}. By examining events with 3 $\alpha$-particles and measuring their relative energies, potential Efimov states were found by looking at 3 $\alpha$-particle triplets where the relative energy between all $\alpha$-particles is consistent with 92 keV. Due to the reaction mechanism used, there is a dominant contribution from uncorrelated $\alpha$-particles which are required to be accounted for via mixing. The remaining spectrum is then accounted for using an arbitrary fit function on top of Breit Wigner peaks which show up around 0.1 MeV suggested mutual $^{8}\mathrm{Be}$ resonances. Due to low statistics and dependence on this fit function, it is therefore difficult to definitely claim evidence of a peak. However, this study represented a dedicated effort to experimentally observe such a state. While in-medium effects afforded by a heavy-ion reaction may enhance the production of an Efimov state, the reaction complexity also introduces numerous sources of background that are difficult to account for. A cleaner population method is through the $\beta$-decay of $^{12}\mathrm{N}/^{12}\mathrm{B}$ which populates $0^{+}$, $1^{+}$, and $2^{+}$ states in accordance to the $\beta$-decay selection rules. Through the recently-developed $\beta$-delayed charged-particle decay technique using the TexAT TPC \cite{NIMJEB}, one is afforded a good probe on any resonance existing at a low relative energy above the 3$\alpha$ threshold. This has not been previously identified due to the difficulties associated with low-energy measurements using implantation in silicon detector arrays. Despite the advantages afforded using this technique, the measurement of a 3-particle final state only 180 keV above threshold is still challenging.

\subsection{Efimov state limits from $\beta$-delayed charged-particle decays in a TPC}
\begin{figure}[h!]
\centerline{\includegraphics[width=0.45\textwidth]{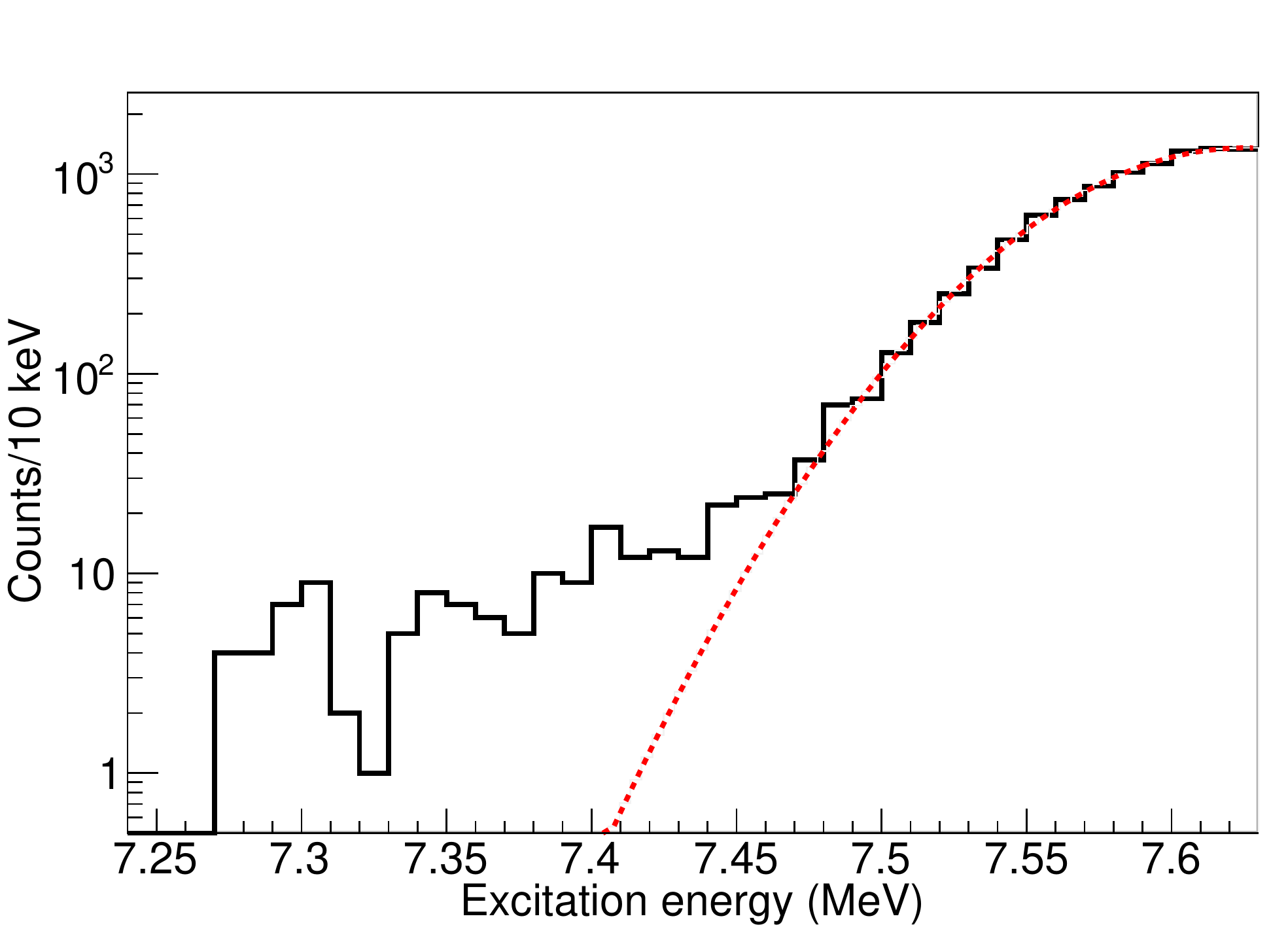}}
\caption{Excitation function obtained from 3-$\alpha$ decay energy deposition. The dotted red line corresponds to a Gaussian fit for the Hoyle state contribution. The yield above this is conservatively taken as possibly arising from additional states such as the Efimov state (ES). The peak around 7.3 MeV corresponds to deposition events on the cathode. \label{fig:ex_low}}
\end{figure}

To establish a limit on the population of the Efimov state via $\beta$-decay of $^{12}\mathrm{N}$, the TexAT TPC was used to study the $\beta$-delayed charged-particle decay of $^{12}\mathrm{N}$ \cite{NIMJEB}. A beam of $^{12}\mathrm{N}$ with an energy of 24 MeV was stopped inside of the active-area of the TexAT TPC by 20 Torr of CO$_2$ gas. This was achieved on a single implantation decay basis which allowed for the matching between the implantation event and the subset of events where the $\beta$-decay populates states above the $\alpha$-threshold in $^{12}\mathrm{C}$. The trigger condition for the decay was that more than one Micromegas pad fired. The trigger threshold is 10 keV and therefore one is sensitive to decay events down to $E_{x} \sim$ 7.4 MeV. The 3D tracks from the 3 $\alpha$-particles arising from these events were then reconstructed and the total energy deposited was used to calculate the excitation energy. This is shown in Fig.~\ref{fig:ex_low} where the majority of events seen are from the Gaussian tails of the Hoyle state at 7.654 MeV. One may also determine the energy of any 3$\alpha$ decay by looking at the total length of the tracks and converting the range of the $\alpha$-particles to an energy and summing. Taking events which lay within the region of interest for both of these plots, one may then manually check events to deduce the origin of these counts. The peak in Fig.~\ref{fig:ex_low} just above the threshold corresponds to events where the implanting beam is scattered (primarily off the entrance window) and the $^{12}\mathrm{N}$ beam is implanted either on the cathode or anode. When it subsequently decays, part of the energy of the event is lost as one or two $\alpha$-particles deposit their energy into the anode or cathode directly rather than liberating electrons in the gas volume. To determine the contribution from states between the absolute threshold at the Hoyle state, a Gaussian tail was fitted for the Hoyle state contribution using the previously obtained experimental resolution of 55 keV. To remain conservative, any background above this Gaussian component was counted as a potential Efimov contribution: 137.1 counts in total. The 95\% confidence limit (C.L.) was therefore taken as 160.5 for $\alpha$-events from Poisson statistics.\par
\begin{figure}[h!]
\centerline{\includegraphics[width=0.5\textwidth]{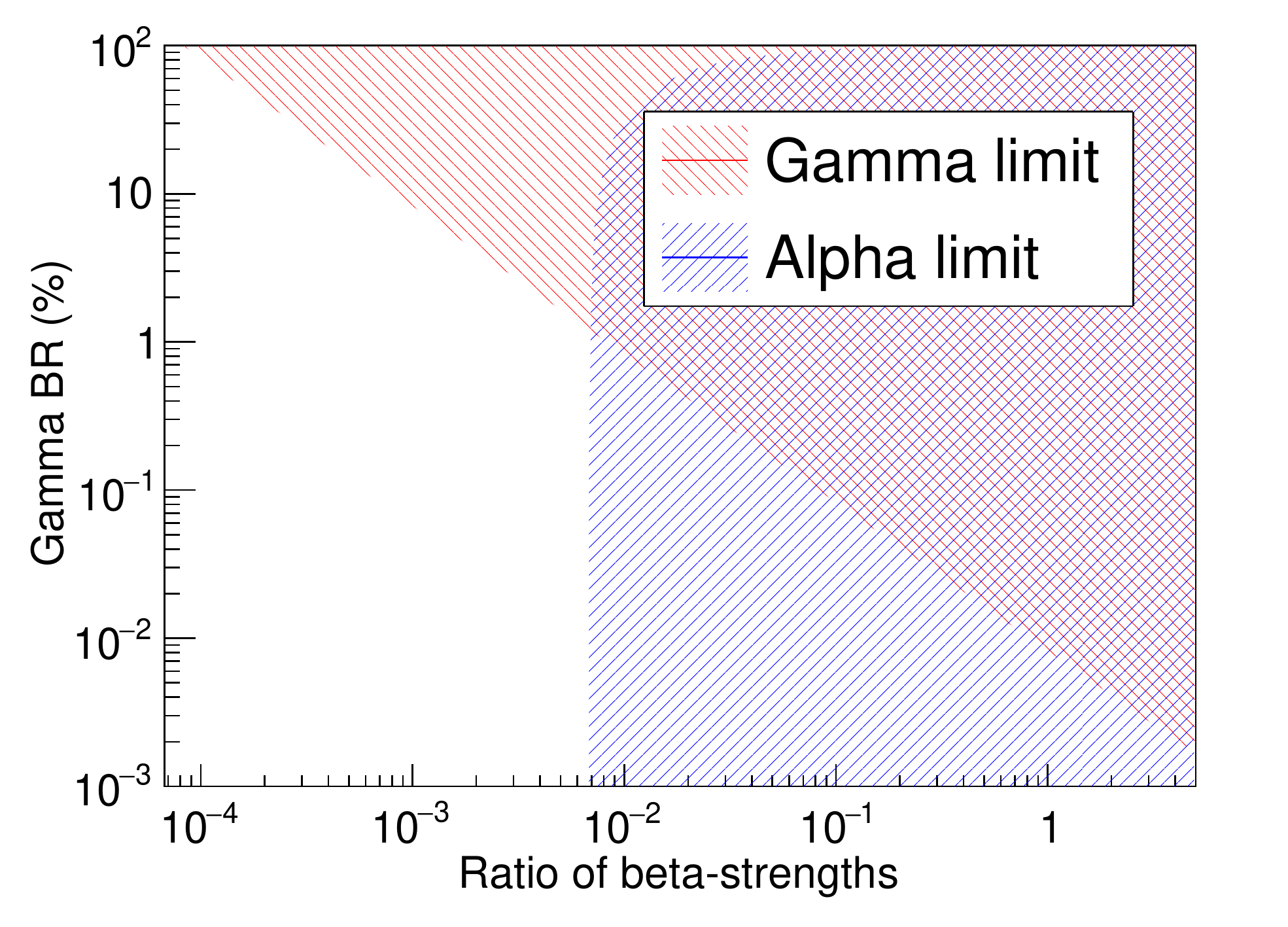}}
\caption{95\% C.L. exclusion plot for a 7.458 MeV Efimov state. The gamma and alpha experimental limits exclude regions of the plot relating to the strength of the $\beta$-feeding strength relative to the Hoyle state (abscissa) and the gamma-ray branching ratio (ordinate). With a particular range of the gamma branching ratio, the existence of a state cannot be conclusively excluded if the $\beta$-feeding strength ratio is $<7 \times 10^{-3}$.  \label{fig:exclusion}}
\end{figure}

To convert this into a limit, the expected number of 3$\alpha$ Efimov-state decays is given by:
\begin{align}
N_{\text{ES}} = N_{\text{Hoyle}} \frac{\beta_{\text{ES}}}{\beta_{\text{Hoyle}}} \frac{\text{BR}_\alpha(\text{ES})}{\text{BR}_\alpha(\text{Hoyle})},
\end{align}
where $\beta_{i}$ describes the $\beta$-decay feeding strength from $^{12}\mathrm{N}$, $\text{BR}_{\alpha}=\frac{\Gamma_{\alpha}}{\Gamma_{\text{tot}}}$ for the ES and the Hoyle state. Re-arranging the terms, one arrives at:
\begin{align}
\frac{\beta_{\text{ES}}}{\beta_{\text{Hoyle}}} \text{BR}_\alpha(\text{ES}) =  \text{BR}_\alpha(\text{Hoyle}) \frac{N_\text{ES}}{N_\text{Hoyle}}. \label{eq:ratio}
\end{align}
The factor on the left-hand side of Eq.~\ref{eq:ratio} allows us to place a limit on the $\alpha$ branching ratio multiplied by the $\beta$-feeding strength relative to the Hoyle state. Inserting the values, $\text{BR}_\alpha(\text{Hoyle}) = 99.9\%$, $N_{\text{Hoyle}} =$ 23,276 and $N_{\text{ES}} < 160.5$. This 95\% confidence limit is represented in Fig.~\ref{fig:exclusion} (slanted-right blue hash region) where $\frac{\beta_{\text{ES}}}{\beta_{\text{Hoyle}}}$ is plotted against $\text{BR}_{\gamma}\text{(ES)} = 1-\text{BR}_\alpha(\text{ES})$. The product of the $\beta$-strengths and the gamma branching-ratio $(\frac{\beta_{\text{ES}}}{\beta_{\text{Hoyle}}} \text{BR}_\alpha(\text{ES}))$ is $<0.69\%$ from Eq.~\ref{eq:ratio} therefore constraining $x(1-y)<0.69\%$ in the exclusion plot. 

\subsection{Efimov state limits from $\beta$-delayed gamma-spectroscopy}
\begin{figure}[h!]
\centerline{\includegraphics[width=0.5\textwidth]{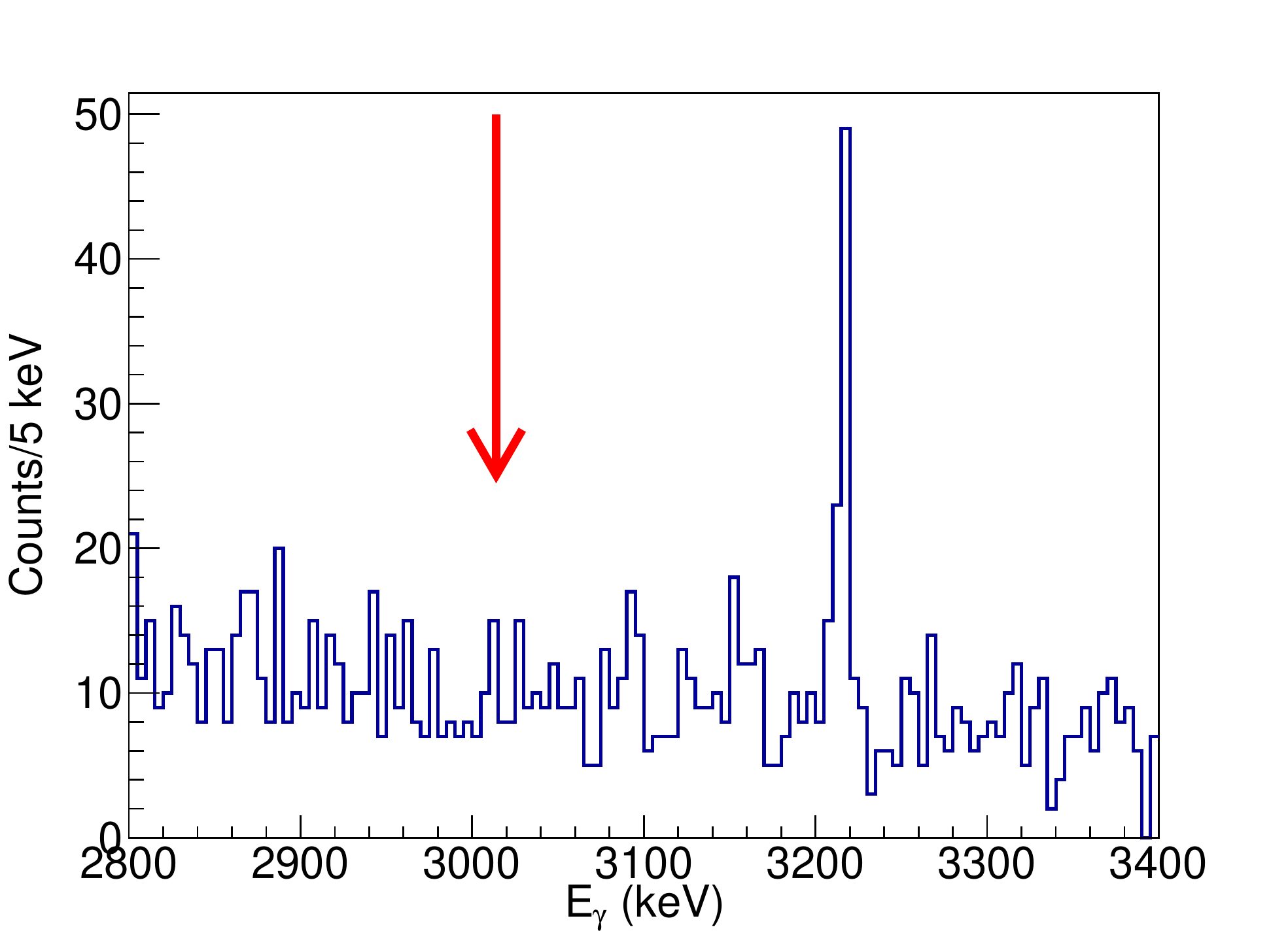}}
\caption{Gammasphere energy spectrum gated on 4439 keV transition. The peak at 3214 keV is from the Hoyle $\rightarrow 2_{1}^{+}$ transition. The yield in the 3014 keV region (red arrow) shows no sign of a peak from the Efimov $\rightarrow 2_{1}^{+}$ transition. Data from \cite{Munch}. \label{fig:Efimov_gammas}}
\end{figure}

Aside from observation in the 3$\alpha$ channel, there have been studies of the $\gamma$-decay spectrum associated with $\beta$-delayed population of $^{12}\mathrm{C}$ using $^{12}\mathrm{B}$ \cite{Munch}. By using Gammasphere to study for coincident gamma rays with the 4.44 MeV first-excited state in $^{12}\mathrm{C}$, one may examine any possible contribution from $^{12}\mathrm{C}$(ES) $\rightarrow {^{12}\mathrm{C}(2_{1}^{+}})$ which should correspond to $E_{\gamma} = (7.458 - 4.444)\, \mathrm{MeV} = 3.014$ MeV. Examining the region of interest from these previous data \cite{Munch}, no such yield can be seen in Fig.~\ref{fig:Efimov_gammas}. The yield from a potential Efimov state is $<11.4$ counts at 95\% C.L. in comparison to the Hoyle state which has a yield of $58~\pm~9$. Following the same prescription above, the number of $\gamma$ rays from Efimov state decays is given by:
\begin{align}
N_{\text{ES}} = N_{\text{Hoyle}} \frac{\beta_{\text{ES}}}{\beta_{\text{Hoyle}}} \frac{\text{BR}_\gamma(\text{ES})}{\text{BR}_\gamma(\text{Hoyle})}.
\end{align}
Separating the unknowns onto the left-hand side from those known on the right-hand side, one gets:
\begin{align}
\frac{\beta_{\text{ES}}}{\beta_{\text{Hoyle}}} \text{BR}_\gamma(\text{ES}) =  \text{BR}_\gamma(\text{Hoyle}) \frac{N_\text{ES}}{N_\text{Hoyle}}.
\end{align}
The known values are $N_{\text{ES}} < 11.4$, $N_{\text{Hoyle}}$ = $58~\pm~9$ and $\text{BR}_\gamma(\text{Hoyle}) =  0.042\% $. This allows for an additional exclusion region from the product of the $\beta$-strengths and the gamma branching-ratio of $<0.008\%$ (red left-slanted hash region in Fig.~\ref{fig:exclusion}). 
\par
The combination of the gamma and alpha measurements in Fig.~\ref{fig:exclusion} demonstrates that an Efimov state at 7.458 MeV is excluded at the 95\% C.L. unless the $\beta$-feeding strength is $<7 \times 10^{-3}$ the strength of the Hoyle state and has a specific decay $\gamma$-decay branch. Requiring the gamma branching ratio to be $\approx 100\%$, this 95\% C.L. extends down to $< 10^{-4}$.\par

The existence of an Efimov state at an energy other than 7.458 MeV may also be examined using the same approach as detailed above. The product of the $\beta$-strengths and the gamma branching-ratio at the 95\% C.L. does not exceed 0.016\%. The technique used for the alpha-limit is robust up until the separation of the Hoyle peak from any potential second peak is roughly 2.35$\sigma$, i.e. up until $E_{x} $= 7.525 MeV where the gamma- and alpha-branching ratios are comparable.
The gamma-limit is highly applicable until the gamma branching ratio is such that the predicted number of Efimov gamma decays is less than the 95\% C.L. or our detection limit. Given that Hoyle state is well measured via gamma-decays in the Gammasphere data and the background is fairly flat around the Hoyle peak, this implies an excitation energy exceeding the Hoyle state is required whereby the gamma branching ratio continues to rapidly decrease, and as such, this scenario may be definitely excluded.

\section{Astrophysical limitations on low-lying states}

\begin{figure}
\centerline{\includegraphics[width=0.5\textwidth]{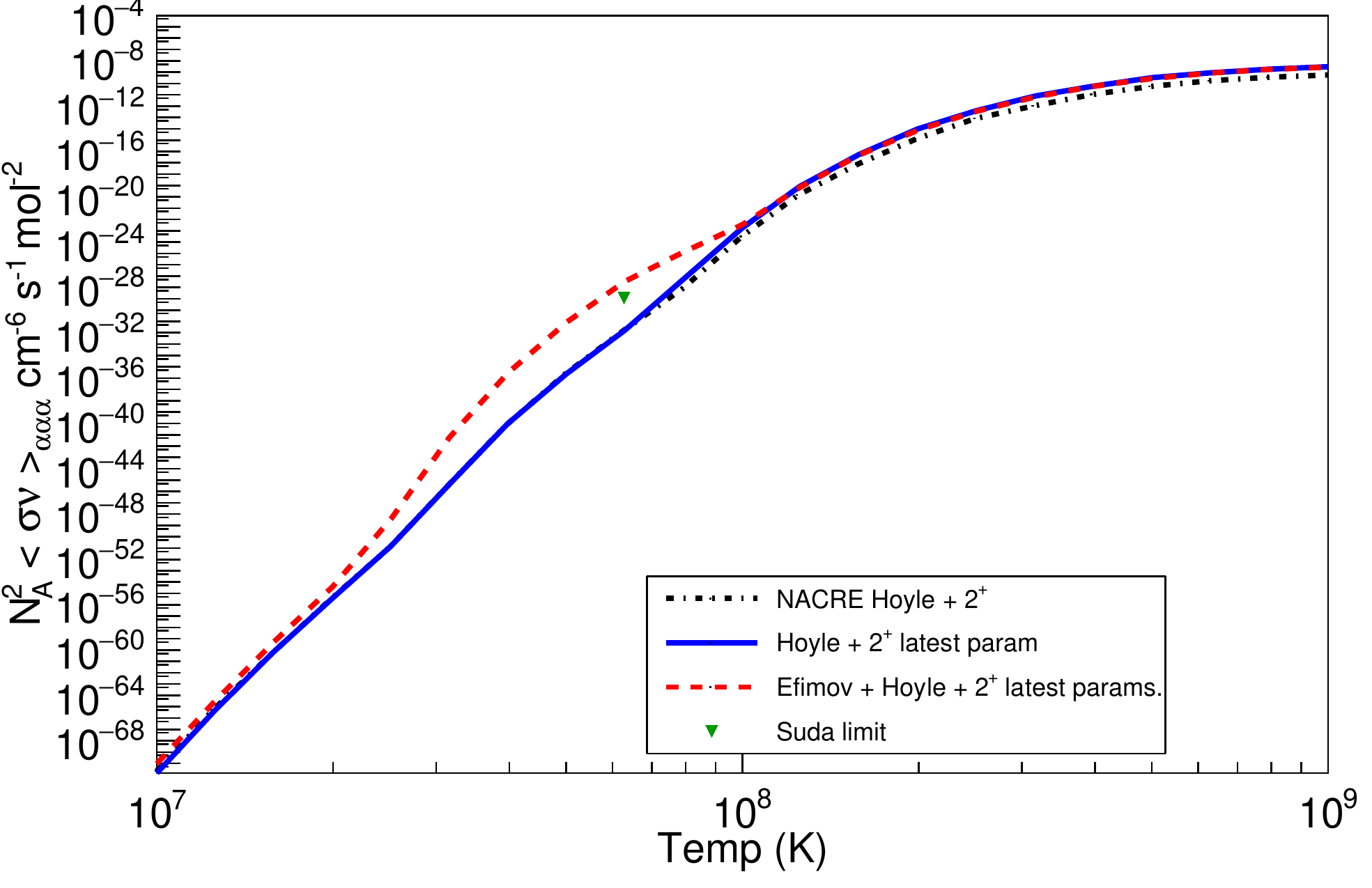}}
\caption{Triple-alpha reaction rates as a function of the astrophysical temperature using the NACRE formulism \cite{NACRE}. The dot-dashed black line corresponds to the pre-existing NACRE rate. The solid blue line includes the Hoyle state and the $2_{2}^{+}$ state in $^{12}\mathrm{C}$ using the latest accepted values. The third line, dashed red, also includes an additional state corresponding to the Efimov state using parameters from Table~\ref{tab:widths}. The upper limit calculated from Suda \cite{Suda} is shown as a green triangle whereby the red giant phase requires a reaction rate of less than $\sim 10^{-29}$ cm$^{-6}$ s$^{-1}$ mol$^{-2}$ at a temperature of $10^{7.8}$ K. The inclusion of the Efimov state clearly violates this reaction rate limit.\label{fig:rates}}
\end{figure}

In the triple-alpha process, there is a strong contribution from low-lying $0^{+}$ states \cite{PhysRevC.87.055804,PhysRevLett.109.141101}. It is for this reason that the Hoyle state so successfully enhances the triple-alpha reaction rate by 7 orders of magnitude, overcoming the A=5, 8 bottleneck in helium-burning stars. Any additional low-lying state therefore must also contribute to a large degree, particularly at lower temperatures. To understand the role that the Efimov state may have, one can examine the expected astrophysical reaction rate including such a resonance in addition to the Hoyle state. The NACRE formulation was used \cite{NACRE} and three cases were examined, the results of which appear in Fig.~\ref{fig:rates}.\par
The first case (dashed-dotted black line) is the NACRE parameter case \cite{NACRE} which shows the current reaction rate. The second case (solid blue line) uses the most up-to-date parameters for the Hoyle state, the $2_{2}^{+}$ in $^{12}\mathrm{C}$ and the $^{8}\mathrm{Be}$ (g.s) \cite{Kibedi,ENSDF,ENSDF2}. The main influence corresponds to the modified radiative and alpha-widths of the Hoyle state and the revised width of the $^{8}\mathrm{Be}$ (g.s). The final case (dashed red line) is whereby, in addition to the states included in the second case, an additional resonance corresponding to the Efimov state was included. For such a state, in Table~\ref{tab:widths}, na\"ive estimates of the widths are calculated via assuming the same underlying reduced width and radiative width for the Efimov state as the Hoyle state when scaled appropriately by their energies (for penetrability and $E_{\gamma}^5$ for the $\alpha$ and $\gamma$ widths accordingly). This is verified by more developed investigations into the properties of an Efimov state \cite{Zhe2018}. For the Efimov state, $\Gamma_{\text{rad}} \gg \Gamma_{\alpha_0}$ so $\Gamma_{\text{rad}} \approx \Gamma_{\text{tot}}$. Therefore, the rate is driven predominantely by $\Gamma_{\alpha_0}$. \par
The contribution of the additional Efimov resonance in the third case clearly demonstrates a phenomenal increase (a factor of $\sim$ 40,000 at $5 \times 10^{7}$ K ) in the reaction rate at lower temperatures which are vitally important for many stars. Understanding the role that the triple-alpha reaction rate has on the dynamics of stars has previously been examined \cite{Suda} and it was concluded that the reaction rate at $10^{7.8}$ K must be less than $\sim 10^{-29}$ cm$^{-6}$ s$^{-1}$ mol$^{-2}$  in order for stars to undergo the red giant phase. This astrophysical limit on the rate is shown in Fig.~\ref{fig:rates} as an upside-down green triangle and the rate from case three (including the Efimov state) can clearly be seen to exceed this limit and does so by a factor of $\sim 24$. As such, the existence of such a state is incompatible with current astrophysical models. Given the contribution of the Efimov state is limited by $\Gamma_{\alpha}$, to quench the contribution of this state to be compatible with the reaction rate limit, this width is required to be decreased by an order of magnitude. The Efimov state has a well-developed cluster structure therefore a partial alpha-width a factor of ten lower than the Hoyle state is clearly not possible. Such a state is additionally excluded down to an extremely small beta-feeding ratio of $10^{-4}$ from the combined $\alpha/\gamma$-spectroscopy for such a state. \par

Modifying the energy of this additional state results in a rate of more than $10^{-29}$ cm$^{-6}$ s$^{-1}$ mol$^{-2}$ for  energies of $7.43 < E_{x} < 7.53$ MeV therefore resonances in this region may also be explicitly excluded.

\section{Conclusion}
Using combined data from particle and gamma-decay from $^{12}\mathrm{N}/^{12}\mathrm{B}$, one may place incredibly restrictive limits on the beta-feeding strength and therefore existence of an Efimov state below the Hoyle state in carbon-12. An Efimov state in such close proximity to the alpha-threshold would decay almost solely via gamma-decay and therefore the beta-feeding strength relative to the Hoyle state is expected to be $<10^{-4}$ at the 95\% C.L. Furthermore, it can clearly be seen that the Efimov state has a tremendous impact on stellar evolution. A low-lying state $0^{+}$ (whether Efimov or merely an $\alpha$-clustered state) would have a significant impact on stellar abundances that the existence of such a resonance is incompatible with current astrophysical models whereby the modified triple-alpha reaction rate would eliminate the red-giant phase of stars.
\begin{table}
\setlength\extrarowheight{2.0pt}
\caption{Na\"ive estimates of the Efimov decay mechanism. The $\alpha$ widths, $\Gamma_{\alpha}$, assume the same reduced width and the radiative width assumes the dominance of the E2 transition to the 4.44 MeV $2^{+}$ state and is scaled as $E_{\gamma}^{5}$.\label{tab:widths}}
\begin{ruledtabular}
\begin{tabular}{ccc}
Parameter&Hoyle state&Efimov state\\ 
\colrule
$E_{x}$&7.654 MeV&7.458 MeV\\
$\Gamma_{\alpha}$ &9.3 eV \cite{ENSDF} &79.8 neV\\ 
$\Gamma_{\mathrm{rad}}$&5.1 meV \cite{Kibedi} &3.72 meV\\
$\Gamma_{\alpha}/\Gamma_{\mathrm{tot}}$&$>99.9\%$&$0.002\%$\\
\end{tabular}
\end{ruledtabular}
\end{table}
\section{Acknowledgements}
J.B. acknowledges and thanks A. Bonasera for discussions on the Efimov state.
This work was supported by the U.S. Department of Energy, Office of Science, Office of Nuclear Science, under award no. DE-FG02-93ER40773 and by National Nuclear Security Administration through the Center for Excellence in Nuclear Training and University Based Research (CENTAUR) under grant number DE-NA0003841. G.V.R. also acknowledges the support of the Nuclear Solutions Institute.

\bibliography{Efimov}

\end{document}